\documentclass[prd,onecolumn,showpacs,nofootinbib,notitlepage,aps,prd,longbibliography]{revtex4-1}
\usepackage{amsmath}
\usepackage{tensor}
\usepackage{latexsym}
\usepackage{amsfonts}
\usepackage{graphicx}
\usepackage{mathrsfs}
\usepackage{hyperref}
\usepackage{mathrsfs}
\usepackage{accents}
\usepackage{color}
\usepackage{xcolor}
\usepackage{subfigure}

\makeatletter
\def\@bibdataout@aps{%
\immediate\write\@bibdataout{%
@CONTROL{%
apsrev41Control%
\longbibliography@sw{%
    ,author="08",editor="1",pages="1",title="0",year="1"%
    }{%
    ,author="08",editor="1",pages="1",title="",year="1"%
    }%
  }%
}%
\if@filesw \immediate \write \@auxout {\string \citation {apsrev41Control}}\fi 
}
\makeatother

\begin{document}

\title{Influence of cosmic expansion on  gravitational waveforms}

\author{Tan Liu$^{1,2}$}\email{lewton@mail.ustc.edu.cn}
\author{Wen-Fan Feng$^{3}$}
\author{Zong-Kuan Guo$^{1,2,4}$}\email{guozk@itp.ac.cn}

\affiliation{$^1$School of Fundamental Physics and Mathematical Sciences, Hangzhou Institute for Advanced Study, UCAS, Hangzhou 310024, China}
\affiliation{$^2$University of Chinese Academy of Sciences, 100049/100190 Beijing, China}
\affiliation{$^3$Kavli Institute for Astronomy and Astrophysics, Peking University, Beijing 100871, China}
\affiliation{$^4$Institute of Theoretical Physics, Chinese Academy of Sciences, Beijing 100190, China}

\begin{abstract}
Gravitational waves undergo redshift as they propagate through the expanding universe, and the redshift may exhibit time-dependent drift. 
Consequently, for any isolated gravitational wave sources, the mass parameter $\mathcal{M}$ and the redshift $z$ exhibit an observational degeneracy, typically manifesting in the waveform as the redshifted mass $\mathcal{M}(1+z)$. Matching together the wave propagation and the wave generation solutions, we show that 
dimensionless source parameters depending on mass $\mathcal{M}$ can break this degeneracy. Notably, the postmerger signal from binary neutron stars contains several dimensionless  parameters that satisfy this condition, including  the quality factors of different frequency components and their frequency ratios. Considering the observations of solely the postmerger signal  by the Neutron star Extreme Matter Observatory  or the Einstein Telescope, based on the Fisher analysis, we find that the redshift  can be measured with fractional uncertainties of $\sim30\%$ for sources at  $0.01<z<0.09$.  Additionally, we present a corrected derivation of the waveform phase correction due to the redshift drift effect, rectifying a sign error  in previous studies.
\end{abstract}

\maketitle

\section{Introduction}
Cosmic expansion can redshift  gravitational waveforms \cite{maggiore2008gravitational}. In the case of accelerating expansion, this redshift exhibits a time-dependent drift \cite{PhysRevLett.87.221103,Yagi_2012}. This means that gravitational waves are stretched (redshifted) by cosmic expansion, and this stretching gradually changes over time (redshift drift). Observing gravitational waves can extract the evolution information of the universe and measure the cosmological parameters \cite{GWHubble,Chen_2021,Wang_2020,Yu_2024}.

\begin{figure}[!htbp]
\begin{center}
\includegraphics[width=8cm]{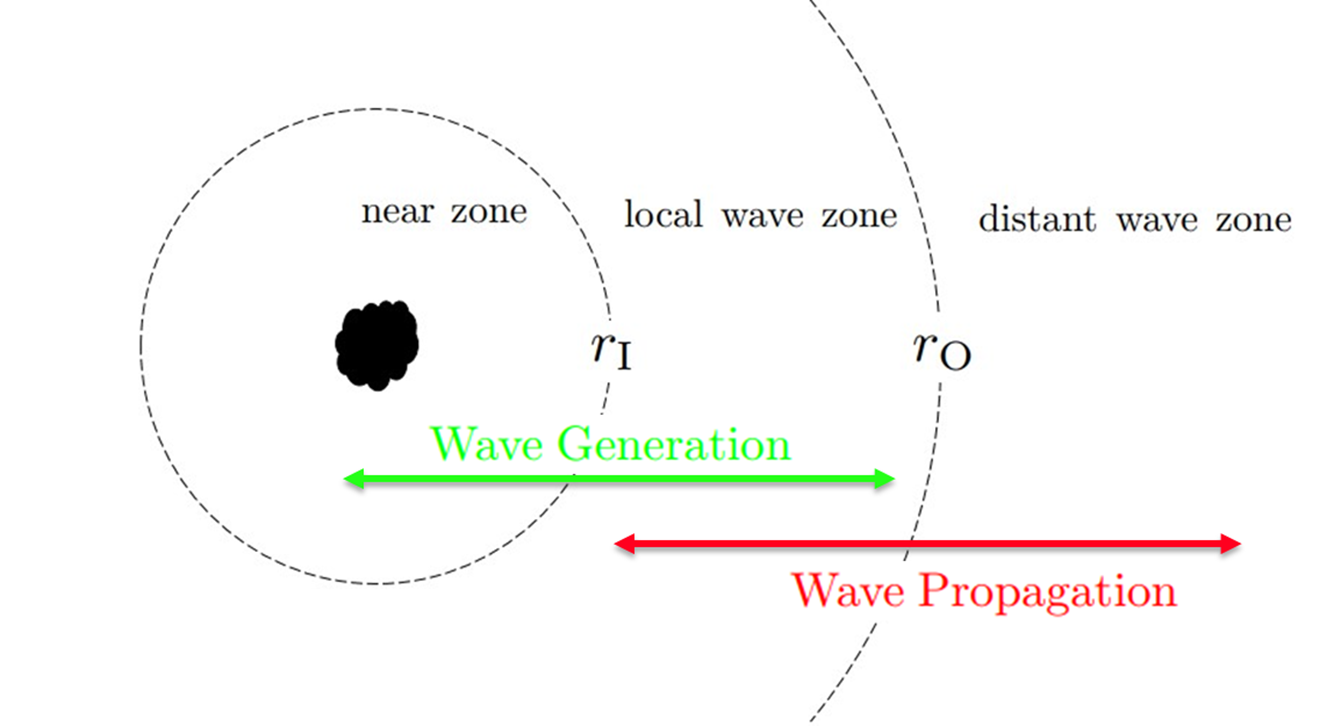}
\caption{Three zones around a source of gravitational waves. The local wave zone is the matching region of the wave generation problem and the wave propagation problem \cite{Thorne2017a}.
}\label{zone}
\end{center}
\end{figure}

To study the generation and propagation of  gravitational waves in an expanding universe, the local wave zone is a crucial concept  \cite{Thorne1987,RevModPhys.52.299}.
In the local wave zone, the gravitational field begins to have wavelike behavior (see Fig. \ref{zone}). This region is large enough that it contains many wavelengths, and small enough that the wave can be approximated to propagate on a flat background. Then, the theory of gravitational waves is split into two parts. First, solve the wave generation problem to obtain  waveforms in the local wave zone. Second, use the waveforms in the local wave zone as the initial condition to solve the wave propagation problem on a curved background. The obtained waveforms in the distant wave zone contain the influence of cosmic expansion. That is, the Einstein field equations should be solved twice and the solutions should take the same form in the overlapping region, the local wave zone. Thorne has used this asymptotic matching technique to study the inspiralling compact binaries and obtained the redshifted inspiralling waveforms \cite{1983grr..proc....1T,Thorne2017a}. For a pedagogical introduction to this method, see Sec. 7.4 in \cite{bender1999advanced}.

The early development of matched asymptotic expansions is closely related to fluid dynamics \cite{Dyke1975,Dong2024}.
Asymptotic matching is introduced in gravitational physics by Fock, see the discussion below Eq. (87.60) in \cite{Fock1964}. This method is widely used in studies of gravitational radiation and motion \cite{PhysRevD.108.024006,Poisson_Will_2014,Blanchet2024,Damour1983,Pound2015,PhysRevD.99.084008,PhysRevD.72.044024,PhysRevD.31.1815,10.1063/1.1724317,PhysRevD.25.2038,PhysRevD.22.1871,PhysRevD.22.1853,Brumberg1989,Brumberg1991,1988CeMec..42..293V}.   Misner \textit{et al.}   (Sec. 20.6 in \cite{misner1973gravitation}) offer an illustrative discussion on motion using this technique.

In this paper, we extend Thorne's work and analyze how cosmic expansion affects gravitational waves from isolated sources. Using matched asymptotic expansions, we derive waveforms that incorporate both redshift and redshift drift effects. The obtained waveform \eqref{wApsi} is valid  in both the local and distant wave zones.
To demonstrate the implementation of matched asymptotic expansions, we work with a toy model: scalar waves in a 1+1 dimensional  universe. The main computation of waves in a four-dimensional universe parallels this model closely.  This model can be solved exactly and contains the redshift effect and the redshift drift effect.

As shown in waveform \eqref{wApsi}, the mass parameter $\mathcal{M}$ and the redshift $z_0$ are degenerate, usually manifesting as the redshifted mass $\mathcal{M}(1+z_0)$.
Breaking the mass-redshift degeneracy could enable gravitational wave sources to serve as more powerful cosmological probes \cite{Mastrogiovanni2024}. Using the waveform \eqref{wApsi}, we demonstrate that the dimensionless parameter depending on mass parameter $\mathcal{M}$ can break this degeneracy. Messenger and Read \cite{PhysRevLett.108.091101} break this degeneracy by utilizing the dimensionless tidal Love number in binary neutron star inspiral waveforms. Subsequently, Messenger \textit{et al.} \cite{PhysRevX.4.041004} show that simultaneously measuring the dimensionless frequency of the postmerger signal and the redshifted mass from the inspiral signal can resolve this degeneracy. We propose extracting the redshift solely from the postmerger signal, as the postmerger waveform contains several mass-dependent dimensionless parameters (e.g., the quality factors
of different frequency components and their frequency ratios \cite{PhysRevD.105.043020,Soultanis2022,Maggiore2018}).
Considering the observations of the postmerger signal by the Neutron star Extreme Matter Observatory (NEMO) \cite{nemo2020,nemopsd} or the Einstein Telescope \cite{Punturo2010,etdpsd}, based on the Fisher analysis \cite{PhysRevD.49.2658}, we find that the redshift  can be measured with fractional uncertainties of $\sim30\%$ for sources at redshifts $0.01-0.09$.
Adding the NEMO to the network of two LIGO A+ observatories can increase the detection rate of the postmerger signal to about one per year \cite{nemo2020}.
Our proposal  enhances the potential for the NEMO to conduct cosmological studies using gravitational wave observations alone \cite{RevModPhys.82.169}.
Additionally, we investigate the redshift drift effect in the inspiralling compact binaries. We rederive the phase correction due to the redshift drift effect and  correct a sign error previous works \cite{PhysRevLett.87.221103,Liu_2024,Yagi_2012,PhysRevD.95.044029}.

In Sec. \ref{sec:FLRW}, we present the procedures to incorporate the redshift effect and the redshift drift effect into the gravitational waveforms in general relativity. In Sec. \ref{sec:applications}, we apply these procedures to postmerger and inspiral waveforms. A highlight is the scenario to break the mass-redshift degeneracy. We conclude and discuss prospects for future research  in Sec. \ref{condis}. In Appendix \ref{toy}, we analyze the scalar waves in the 1+1 dimensional universe  as a pedagogical example for the method of asymptotic matching.

For the metric, Riemann, and Ricci tensors, we follow the conventions of Misner \textit{et al.} \cite{misner1973gravitation}. We adopt the unit $G_N=c=1$, with $c$ being the speed of light and $G_N$ being the gravitational constant. 
We have used the cosmological parameters $H_0=67.66~\mathrm{km/s/Mpc}$, $\Omega_m=0.311$, and $\Omega_\Lambda=1-\Omega_m$ \cite{2020A&A...641A...6P} to compute the luminosity distance $D_L$.








\section{Influence of cosmic evolution on gravitational waveforms in general relativity}\label{sec:FLRW}
We now use the method of matched asymptotic expansions to study the influence of cosmic expansion on gravitational waveforms. Three zones are introduced around the wave source, see Fig. \ref{zone}. If $r$ is the coordinate distance from the wave source, then the near zone is defined by $r\lesssim r_{\text{I}}$, with the inner radius $r_{\text{I}}$ marking the boundary of the zone. The inner radius $r_{\text{I}}$ should be  larger than the wavelength and the gravitational field begins to exhibit wave behavior there. The local wave zone is defined by $r_{\text{I}}\lesssim r \lesssim r_{\text{O}}$. The outer radius $r_{\text{O}}$ should be much larger than the wavelength, so that the local wave zone contains many wavelengths, but much smaller than the curvature radius of the background universe. The distant wave zone is given by $r \gtrsim r_{\text{O}}$. We can then deal with the wave generation problem on a flat background $(r \lesssim r_{\text{O}})$ and the wave propagation problem on a curved background $(r \gtrsim r_{\text{I}})$. The local wave zone is the matching region and the solutions to the wave generation problem and the wave propagation problem should be equal in this region. For a more delicate separation of regions, see \cite{Thorne2017a} .

\subsection{Solution of the wave generation problem in the local wave zone}
In the local wave zone, in harmonic coordinates $(t,x,y,z)$, the dominant part of the metric $g_{\mu\nu}$, for which the expansion along the null ray, $r\equiv\sqrt{x^2+y^2+z^2}\to +\infty$ with fixed retarded time
\begin{equation}
u\equiv t-r,
\end{equation}
takes the form
\begin{equation}\label{general_local}
g_{\mu\nu}=\eta_{\mu\nu}+\frac{K_{\mu\nu}(u,\theta,\phi)}{r}+\mathcal{O}(\frac{1}{r^2}).
\end{equation}
Here, $\eta_{\mu\nu}$ is the Minkowski metric and $K_{\mu\nu}$ are computable functions of the source multipole moments \cite{Blanchet2024}. The angles are given by $x=r \sin\theta \cos\phi$ and $y=r\sin\theta\sin\phi$. Contracting the polarization tensors with the $\mathcal{O}(\frac{1}{r})$ part of the metric yields the `plus' and `cross' polarization waveforms \cite{Thorne2017a,Blanchet2024}
\begin{equation}\label{local wave}
h_+=\Re[A_+e^{i\psi_+}] \text{ with}\quad A_+ = \frac{p^{(L)}(u,\theta,\phi)}{r},\quad \psi_+=\psi_+(u);
\end{equation}
\begin{equation}
h_\times=\Re[A_\times e^{i\psi_\times}] \text{ with} \quad A_\times = \frac{c^{(L)}(u,\theta,\phi)}{r},\quad \psi_\times=\psi_\times(u).
\end{equation}
The above result applies to isolated gravitational wave sources. The superscript $(L)$ denotes that the  result only apply in the local wave zone, since the coordinate $(t,r,\theta,\phi)$ cover only a small region of the universe. For compact binary systems, both the inspiral waveforms (Eqs. (27.71) in \cite{Thorne2017a}) and the ringdown waveforms (Eq. (14.102) in \cite{Maggiore2018}) take this form\footnote{Actually, in the harmonics coordinates, there are terms taking the form $\frac{\ln(r)}{r}$, see Eq. (11.297e) in \cite{Poisson_Will_2014} and Eq. (87.59) in \cite{Fock1964}. In the multipolar post-Minkowskian formalism, logarithms only arise due to the integral of the non-compact source terms that behave like $\frac{1}{r^2}$. The appearance of logarithms is a coordinate effect and they disappear in Bondi-type coordinates \cite{Blanchet2024}. However, the Bondi-Sachs formalism is not well developed in the distant wave zone \cite{PhysRevD.108.064039,PhysRevD.102.104043}. The examples studied in Sec. \ref{sec:applications} do not contain any logarithms.  Therefore, we will ignore the logarithms in this paper and leave these for future work.}.

{The amplitude and phase depend on the source parameters. Using the   Pi theorem \cite{PhysRev.4.345}, this dependence can be demonstrated as follows}
\begin{equation}\label{localApsi}
A_+=\frac{\mathcal{M}}{r} Q^{(L)}(\frac{u-u_0}{\mathcal{M}},\theta,\phi,\Pi), \quad \psi_+=\psi_+^{(L)}(\frac{u-u_0}{\mathcal{M}},\Pi).
\end{equation}
Here, $\mathcal{M}$ is the typical mass of the source. For a binary system, $\mathcal{M}$ may represent any of the following: the chirp mass, the total mass, an individual component mass, or other related quantities. For a rotating neutron star, $\mathcal{M}$ can be the mass of the neutron star. The  two functions $Q^{(L)}$ and $\psi_+^{(L)}$ and their arguments are all dimensionless. This ensures that the amplitude $A_+$ is dimensionless. $\Pi$ denotes the dimensionless source parameters collectively. For a binary neutron star system, the mass ratio and the dimensionless tidal Love number belong to $\Pi$. The reference retarded time $u_0$ is introduced for later convenience. The above result is the solution to the wave generation problem.  The cross polarization can be expressed in the same way. See Eq. (5.260) in \cite{maggiore2008gravitational} for the gravitational wave phase.

\subsection{Propagation of the waves in the local and distant wave zones}
We now use geometric optics in the local and distant wave zones to propagate the waveforms and obtain the waveforms in the distant wave zone. The gravitational wave $h_{\mu\nu}$ can be viewed as a perturbation around the  Friedmann-Lema\^{i}tre-Robertson-Walker (FLRW) metric $\bar{g}_{\mu\nu}$
\begin{equation}
g_{\mu\nu}=\bar{g}_{\mu\nu}+h_{\mu\nu}.
\end{equation}
The FLRW metric takes the form
\begin{equation}\label{flrw}
\bar{g}_{\mu\nu}dx^\mu dx^\nu =a^2(\eta)[-d\eta^2+d\chi^2+\Sigma^2(d\theta^2+\sin^2\theta ~d\phi^2)]
\end{equation}
where $\Sigma \equiv \chi \text{ for a spatially flat universe}, \sin\chi \text{ for a closed universe}, \sinh\chi \text{ for an open universe}$, and $\eta$ is the conformal time. Note that the argument of this subsection applies to arbitrary scale factor $a(\eta)$ and arbitrary spatial curvature. 

In the local wave zone, when the time duration is much smaller than the age of the universe, the scale factor is approximated to be a constant and the FLRW metric reduces to the flat metric. We have 
\begin{equation}
t=a\eta, \qquad r=a\Sigma.
\end{equation}

Expanding the Einstein field equations around the FLRW background yields the linear perturbation equation \cite{maggiore2008gravitational}
\begin{equation}
-\frac12\bar{\square}\hat{h}_{\mu\nu}+\tensor{\hat{h}}{_{\beta(\mu|}^\beta_\nu_)}-\frac12\bar{g}_{\mu\nu}\tensor{\hat{h}}{_{\alpha\beta|}^{\alpha\beta}}=0
\end{equation}
Here $\bar{\square}$ is given by $\bar{\nabla}_\mu\bar{\nabla}^\mu$, and both $\bar{\nabla}_\mu$ and the slash $|$ denote the covariant derivative such that $\bar{\nabla}_\mu\bar{g}_{\alpha\beta}=0$. The background metric $\bar{g}^{\mu\nu}$ is used to raise the indices. The hat indicates the trace-reversed part of a tensor, e.g., $\hat{h}_{\mu\nu}=h_{\mu\nu}-\frac12\bar{g}^{\mu\nu}\bar{g}^{\alpha\beta}h_{\alpha\beta}$.
Indices placed between round brackets are symmetrized. The terms with no derivatives acting on $\hat{h}_{\mu\nu}$ are discarded since we focus on oscillating fields.

Using the harmonic gauge condition
\begin{equation}
\bar{\nabla}^{\mu}\hat{h}_{\mu\nu}=0
\end{equation}
and the geometric optics approximation, we obtain, in the local and distant wave zones, the phase evolution equation \cite{maggiore2008gravitational}
\begin{equation}
k^\alpha k_\alpha =0,
\end{equation}
and the amplitude evolution equation \cite{maggiore2008gravitational}
\begin{equation}
2k^\alpha {\bar{\nabla}}_\alpha A_+ + A_+ {\bar{\nabla}}^\alpha k_\alpha=0,
\end{equation}
where the wave vector $k^\alpha$ is given by
\begin{equation}
k^\alpha = -\bar{g}^{\alpha\beta}\partial_\beta \psi_+=\frac{dx^\alpha}{dl}
\end{equation}
with $l$ the affine parameter along the null ray. The above two evolution equations are similar to that of the scalar wave toy model in Appendix \ref{toy}.
Since the evolution of the cross section of the bundle of the null rays is determined by \cite{Thorne2017a,PhysRevD.108.024006} 
\begin{equation}\label{nablak}
k^\mu\bar{\nabla}_\mu \ln(a^2\Sigma^2)=\bar{\nabla}^\alpha k_\alpha,
\end{equation}
the evolution equations of the amplitude and phase along a null ray become (cf. Fig. \ref{ray})
\begin{equation}
\frac{d}{dl}(A_+ a\Sigma)=0, \quad \frac{d\psi_+}{dl}=0.
\end{equation}
The solutions to the above equations are
\begin{equation}
A_+=\frac{p^{(W)}(t_r,\theta,\phi)}{a\Sigma},\quad \psi_+=\psi_+^{(W)}(t_r,\theta,\phi).
\end{equation}
The superscript $(W)$, where $W$ stands for `wave', denotes that the results apply in the local and distant wave zones. Here, the retarded time is given by \cite{Thorne2017a}
\begin{equation}\label{tr}
t_r = \int_0^{\eta - \chi} a(\eta') d\eta'
\end{equation}
which is similar to the retarded time in Appendix \ref{toy} and reduces to $t_r=u$ in the local wave zone. Given a set of $(t_r, \theta,\phi)$, it corresponds to a null ray and the functions  $p^{(W)}$  and $\psi_+^{(W)}$ are constant along this null ray. These functions are to be determined by waveforms in the local wave zone. In the local wave zone, the amplitude and the phase should reduce to Eq. \eqref{local wave} and satisfy the matching condition along a null ray
\begin{equation}
\frac{p^{(W)}}{a\Sigma}=\frac{p^{(L)}}{r},\quad \psi_+^{(W)}=\psi_+^{(L)}, \quad \text{~for~} \Sigma\to 0 \text{~with~} t_r,~ \theta,~\phi \text{~fixed}
\end{equation}
which yields
\begin{equation}\label{amp-pha}
p^{(W)}=\mathcal{M} Q^{(L)}(\frac{t_r-t_{r0}}{\mathcal{M}},\theta,\phi,\Pi),\qquad \psi_+^{(W)}=\psi_+^{(L)}(\frac{t_r-t_{r0}}{\mathcal{M}},\Pi).
\end{equation}
Here, the retarded time $t_{r0}$ corresponds to $u_0$ in Eq. \eqref{localApsi}. For ease of application, the retarded time $t_r$ should be expressed in terms of the observed physical time $t=\int_0^{\eta} a(\eta') d\eta'$.
Using Eq. \eqref{tr}, the relation between the retarded time $t_r$ and the physical time $t$ is \cite{PhysRevLett.87.221103}
\begin{equation}
    \int^t_{t_r} \frac{dt'}{a(t')}=\chi. 
\end{equation}
Fixing $\chi$, and differentiating the above equation with respect to $t_r$ yields \cite{PhysRevLett.87.221103} 
\begin{equation}
    \frac{dt}{dt_r}=1+z, \quad \frac{d^2t}{dt_r^2}=(1+z)[(1+z)H_0-H(z)]\equiv g(z),
\end{equation}
where $H_0$ and $H(z)$ are respectively the Hubble parameters at present and at redshift $z$. The second equation represents the  redshift drift (or cosmic acceleration) effect.

If we ignore the redshift drift effect, which amounts to ignoring the time evolution of observed frequency of the scalar waves in Fig. \ref{omega}, the physical time can be approximated by 
\begin{equation}
t-t_0=(t_r-t_{r0})(1+z_0),
\end{equation}
where $t_0$ is the observed physical time corresponding to $t_{r0}$ and $z_0$ is the source redshift at time $t_{r0}$.
Then, the amplitude and the phase can be rewritten as
\begin{equation}\label{redwave}
A_+=\frac{\mathcal{M}(1+z_0) }{D_L}Q^{(L)}(\frac{t-t_0}{\mathcal{M}(1+z_0)},\theta,\phi,\Pi),\qquad \psi_+=\psi_+^{(L)}(\frac{t-t_0}{\mathcal{M}(1+z_0)},\Pi).
\end{equation}
Here, $D_L=a\Sigma (1+z_0)$ is the luminosity distance.
Thus, we obtain the waveform which is applicable in the local and distant wave zones and incorporates the redshift effect.

Considering the redshift drift effect, the physical time can be approximated by  \cite{PhysRevLett.87.221103} 
\begin{equation}
t-t_0=(t_r-t_{r0})(1+z_0)+\frac12 g(z_0)(t_r-t_{r0})^2.
\end{equation}
Keeping the first order term of $g(z_0)$, this equation can be rewritten as
\begin{equation}\label{driftt}
    t_r-t_{r0}=\frac{1}{1+z_0}(\Delta t-X(z_0)\Delta t^2)
\end{equation}
with $\Delta t\equiv t-t_0$. The acceleration parameter $X$ is defined as  \cite{PhysRevLett.87.221103} 
\begin{equation}
    X(z)\equiv \frac12 \frac{g(z)}{(1+z)^2}.
\end{equation}
Substituting Eq. \eqref{driftt} into Eq. \eqref{amp-pha}, we obtain the waveform 
\begin{equation}\label{wApsi}
A_+=\frac{\mathcal{M}(1+z_0)}{D_L}Q^{(L)}(\frac{\Delta t-X(z_0)\Delta t^2}{\mathcal{M}(1+z_0)},\theta,\phi,\Pi),\qquad \psi_+=\psi_+^{(L)}(\frac{\Delta t-X(z_0)\Delta t^2}{\mathcal{M}(1+z_0)},\Pi),
\end{equation}
which incorporates both the redshift effect and the redshift drift effect. 
The relation between the observed gravitational wave frequency $\omega_o$ and the emission frequency $\omega_e$ is
\begin{equation}\label{omegao_x}
    \omega_o=\frac{d\psi_+}{dt}= \frac{d\psi_+}{dt_r}\frac{dt_r}{dt}=\omega_e\frac{1}{1+z_0}[1-2X(z_0)(t-t_0)].
\end{equation}
If the expansion of the universe is accelerating, $X(z_0)>0$, and the emission frequency is constant, then the observed frequency will decrease with time\footnote{In the flat $\Lambda$CDM model, $X(z)>0$ for $0<z<1.96$. Therefore, this statement only applies for sources at $0<z<1.96$. }. This is consistent with cases II and III of the scalar waves in Fig. \ref{omega}.

In summary, from the local wave zone waveform \eqref{localApsi} we can obtain  the waveform \eqref{wApsi}, which is valid in both the local and distant wave zones, by the following procedures. The source distance $r$ in Eq. \eqref{localApsi}  should be replaced by the luminosity distance $D_L$, the retarded time difference $u-u_0$ should be replaced by $\Delta t-X(z_0)\Delta t^2$, the mass parameter $\mathcal{M}$ should be replaced by the redshifted mass parameter $\mathcal{M}(1+z_0)$, and the dimensionless quantities in $\Pi$ are not affected. If there is a parameter of dimension $[\text{mass}]^n$, we  divide it by $\mathcal{M}^n$ to form a dimensionless parameter. To ensure that the dimensionless parameter formed is not affected, the  parameter of dimension $[\text{mass}]^n$ should be multiplied by $(1+z_0)^n$. These procedures also apply to the cross polarization.













\section{Astrophysical applications of the redshift effect and the redshift drift effect}\label{sec:applications}

\subsection{Breaking the mass-redshift degeneracy}
In this subsection, we will focus on the redshift effect and ignore the redshift drift effect. It can be seen from Eq. \eqref{redwave} that the mass and the redshift usually appear in the combination redshifted mass $\mathcal{M}(1+z_0)$. If there are dimensionless parameters in $\Pi$ which depend on the mass $\mathcal{M}$, then it is possible to break the mass-redshift degeneracy. Messenger \textit{et al.}  find that using the tidal effect in the inspiral signal of binary neutron stars \cite{PhysRevLett.108.091101}, or the postmerger signal of binary neutron stars together with the inspiral signal \cite{PhysRevX.4.041004}, can break this degeneracy. 
We will show that the postmerger signal alone can break this degeneracy, after briefly introducing the idea of Messenger and Read \cite{PhysRevLett.108.091101}.

In the local wave zone, the Newtonian order waveform of the plus polarization emitted by an inspiralling neutron star binary on a quasicircular orbit with  masses $m_1$ and $m_2$ is (Sec. 4.1 in \cite{maggiore2008gravitational}) 
\begin{equation}\label{newhp}
    h_+(u)=\frac{4 {M_c}}{r} \left(\frac{5M_c}{256(u_c-u)}\right)^{1/4} \frac{1+\cos^2\theta}{2}\cos[2\Phi(u-u_c, M_c)].
\end{equation}
Here $M_c\equiv \frac{(m_1 m_2)^{3/5}}{(m_1+m_2)^{1/5}}$ is the chirp mass  and  the 
reference retarded time is chosen to be the coalescence time in the source frame  $u_c$. The orbital phase is \cite{maggiore2008gravitational}
\begin{equation}
    \Phi(u-u_c, M_c)=-2\left(\frac{u_c-u}{M_c}\right)^{5/8}+\Phi_c
\end{equation}
with $\Phi_c$ the phase at coalescence.
The mass parameters $m_1$ and $m_2$ and the orbital phase $\Phi$ are defined in the source frame.

Using Eq. \eqref{redwave}, the waveform incorporating the redshift effect becomes
\begin{equation}
    h_+(t)=\frac{4 {\bar{M}_{c}}}{D_L} \left(\frac{5\bar{M}_{c}}{256(t_c-t)}\right)^{1/4} \frac{1+\cos^2\theta}{2}\cos[2\Phi(t-t_c, \bar{M}_{c})],
\end{equation}
where $\bar{M}_{c}\equiv M_c(1+z_c)$ is the redshifted chirp mass with $z_c$ the source redshift at coalescence. $t_c$ is the observed coalescence time.
It can be seen that the redshift $z_c$ is degenerate with the source frame chirp mass $M_c$.

The tidal interaction between the neutron stars can influence the orbital evolution of the binary system. Considering the adiabatic tidal effect, the dimensionless tidal deformabilities\footnote{$\Lambda_1$ corresponds to $\frac{\lambda_1}{m_1^5}$ in \cite{PhysRevLett.108.091101}.} of neutron stars, $\Lambda_1$ and $\Lambda_2$, will enter into the phase $\Phi$. Given the equation of state of the neutron star, the dimensionless tidal deformability is a function of the source frame neutron star mass $\Lambda_i(m_i)~ (i=1,2)$. By measuring the dimensionless parameters $\Lambda_1$ and $\Lambda_2$, the source frame masses can be provided, and together with the redshifted chirp mass $\bar{M}_{c}$, the redshift $z_c$ can be determined.
This is the proposal of Messenger and Read \cite{PhysRevLett.108.091101}.

We will now demonstrate that the mass-redshift degeneracy can also be broken by solely leveraging the postmerger signal.
Informed by numerical relativity simulations, the postmerger signal contains four exponentially decaying sinusoids \cite{PhysRevD.105.043020} 
\begin{eqnarray} 
\label{analytic postmerger}
h_\mathrm{+}(u) &=&\sqrt{\frac{5}{4\pi}}\frac{M}{r} \mathcal{N} \Big[A_\mathrm{peak}\  e^{(-u/\tau_\mathrm{peak})}\cdot\sin(\phi_\mathrm{peak}(u))\nonumber\\ 
&+& A_\mathrm{spiral} \  e^{(-u/\tau_\mathrm{spiral})}\cdot\sin(2\pi f_\mathrm{spiral}\cdot u+\phi_\mathrm{spiral})\nonumber\\
&+& A_{2-0}\  e^{(-u/\tau_{2-0})}\cdot \sin(2\pi f_{2-0}\cdot u+\phi_{2-0})\nonumber\\
&+& A_{2+0}\  e^{(-u/\tau_{2+0})}\cdot \sin(2\pi f_{2+0}\cdot u+\phi_{2+0})\Big], 
\end{eqnarray}
where the $f_\mathrm{peak}$ component's phase, $\phi_\mathrm{peak}(u)$, is \cite{PhysRevD.105.043020} 
\begin{eqnarray}\label{phipeak(u)}
\phi_\mathrm{peak}(u) =    \left\{
\begin{array}{ll}
2\pi \left(f_\mathrm{peak,0}+\frac{\zeta_\mathrm{drift}}{2}u \right) u+\phi_\mathrm{peak}, & \mbox{for } u\leq u_* \\
2\pi \left(f_\mathrm{peak,0}+ \zeta_\mathrm{drift} u_* \right)(u-u_*)+ \phi_\mathrm{peak}(u_*). &\mbox{for } u> u_*
\end{array} 
\right.\quad 
\end{eqnarray}
The dimensionless normalization factor $\mathcal{N}$ is introduced to better fit  the numerical relativity simulations. The reference retarded time is  $u_0=0$ corresponding to the merging time. $\phi_\mathrm{peak}=\phi_\mathrm{peak}(0)$ is the initial phase of the $f_\mathrm{peak}$ component.
The $f_\mathrm{peak}$ component is dominant over the other three components.  The frequency $f_\mathrm{peak}$ initially drifts linearly with time and then becomes a constant \cite{PhysRevD.105.043020}. For simplicity, the orientation of the binary system is set to be face-on, so there is no angular dependence in the above local wave zone waveform.

Soultanis \textit{et al.} \cite{PhysRevD.105.043020} use the equation of state MPA1 \cite{MUTHER1987469} which is compatible with the constraints from GW170817 and PSR J0348+0432.
They simulate equal mass neutron star binary systems, initially on a circular orbit, and find that for source frame total masses $M$ in the range $2.4-3.1~ M_\odot$,  there exists the postmerger signal and  no black hole is formed immediately after merging. For different total masses, the dimensionless amplitudes $A_k$, the frequencies $f_k$, the decay timescales $\tau_k$, the initial phases $\phi_k$  (for $k$ = peak, spiral, $2\pm 0$), and the normalization factor $\mathcal{N}$ can be extracted from the numerical relativity simulations. These parameters are functions of the source frame total mass\footnote{We expect that more comprehensive simulations in the future will  show that these model parameters also depend on other parameters of the binary neutron star system, such as the mass ratio, the spin parameters, and the orbital eccentricity. Nevertheless, this simplified postmerger waveform model is sufficient to show that the postmerger signal alone can break the mass-redshift degeneracy. } $M$ and are modelled by polynomials of $M$. The fitting results of the dominant $f_\mathrm{peak}$ component are \cite{PhysRevD.105.043020}
\begin{eqnarray}
\label{equations-fpeak(t)-fits_1}
\zeta_\mathrm{drift} &=& -1.420 ~M^3+11.085 ~ M^2- 28.834~M+24.943,\\
\label{equations-fpeak(t)-fits_2}f_\mathrm{peak,0} &=& +0.908~M^2-3.974~M+7.058,\\
\label{equations-fpeak(t)-fits_3}u_{*} &=& -8.523~M^2+40.179~M-40.741,
\end{eqnarray}

\begin{eqnarray}
 \label{Apeak of mtot}A_\mathrm{peak} &=& -0.409~M^2+3.657~M-6.130, \\
 \label{Tpeak of mtot}\tau_\mathrm{peak} &=& +7.782~M^2-53.040~M+93.542,
\end{eqnarray}

\begin{eqnarray}
\label{phipeak-Mtot}\phi_{\mathrm{peak}} =    \left\{
\begin{array}{ll}
+18.957~M-46.321,  & \mbox{for } M\leq 2.7~M_\odot \\
+43.425~M-113.152. &\mbox{for }  M >  2.7~M_\odot  \\
\end{array} 
\right.
\end{eqnarray}
The three parameters $\zeta_\mathrm{drift}$, $f_\mathrm{peak,0}$ and $u_{*}$ determine the time evolution of $f_\mathrm{peak}$. The frequency $f_\mathrm{peak}$ initially decreases linearly from  $f_\mathrm{peak,0}$  with the rate  $\zeta_\mathrm{drift}$, then becomes constant after $u_{*}$. 
The fitting formulae of the three secondary components are collected in Appendix \ref{sec:fitting_secondary}. Table \ref{Units} provides the units of the parameters in the  fitting formulae.

Using Eq. \eqref{redwave} to incorporate the redshift effect into the postmerger waveform \eqref{analytic postmerger}, we obtain
\begin{eqnarray} 
\label{red postmerger}
h_\mathrm{+}(t) &=&\sqrt{\frac{5}{4\pi}}\frac{\bar{M}}{D_L} \mathcal{N} \Big[A_\mathrm{peak}\  e^{(-t/\bar{\tau}_\mathrm{peak})}\cdot\sin(\phi_\mathrm{peak}(\frac{t}{1+z_0}))\nonumber\\ 
&+& A_\mathrm{spiral} \  e^{(-t/\bar{\tau}_\mathrm{spiral})}\cdot\sin(2\pi \bar{f}_\mathrm{spiral}\cdot t+\phi_\mathrm{spiral})\nonumber\\
&+& A_{2-0}\  e^{(-t/\bar{\tau}_{2-0})}\cdot \sin(2\pi \bar{f}_{2-0}\cdot t+\phi_{2-0})\nonumber\\
&+& A_{2+0}\  e^{(-t/\bar{\tau}_{2+0})}\cdot \sin(2\pi \bar{f}_{2+0}\cdot t+\phi_{2+0})\Big], 
\end{eqnarray}
where $z_0$ is the source redshift at the merging time and $D_L$ is the luminosity distance of the binary neutron stars. The overhead bar  indicates that the corresponding quantity is redshifted.

Through observing the postmerger wave, one can measure the redshifted initial peak frequency $\bar{f}_\mathrm{peak,0}$ and the redshifted decay timescale $\bar{\tau}_\mathrm{peak}$.
Multiplying these two quantities yields a dimensionless parameter $\bar{f}_\mathrm{peak,0}\cdot\bar{\tau}_\mathrm{peak}={f}_\mathrm{peak,0}\cdot{\tau}_\mathrm{peak}$ which is not influenced by the redshift effect. It can be seen from Eqs. \eqref{equations-fpeak(t)-fits_2} and \eqref{Tpeak of mtot} that the product ${f}_\mathrm{peak,0}\cdot{\tau}_\mathrm{peak}$ is determined by the source frame total mass $M$. That is, by measuring the redshifted frequency $\bar{f}_\mathrm{peak,0}$ and the redshifted timescale  $\bar{\tau}_\mathrm{peak}$, the source frame mass $M$ can be inferred, from which  the source frame frequency ${f}_\mathrm{peak,0}$ and timescale ${\tau}_\mathrm{peak}$ can be derived. In this way, the redshift $z_0$ can be extracted.  

There are other observable dimensionless  parameters, determined by the source frame mass $M$, which can be used to infer the redshift $z_0$, such as the product $\bar{\zeta}_\mathrm{drift}\cdot \bar{\tau}^2_\mathrm{peak}$, the product $\bar{\tau}_\mathrm{peak}\cdot\bar{f}_\mathrm{spiral}$, the frequency ratio $\bar{f}_\mathrm{peak,0}/\bar{f}_\mathrm{spiral}$, and the timescale ratio $\bar{\tau}_\mathrm{peak}/\bar{\tau}_\mathrm{spiral}$\footnote{We do not use the dimensionless amplitudes $A_k$ and initial phases $\phi_k$ for  $k$ = peak, spiral, $2\pm 0$ to infer the redshift $z_0$, since future more comprehensive simulations will demonstrate that the amplitudes and phases may have complicated dependence on other source parameters besides the total mass $M$ \cite{PhysRevLett.130.021001}. }.

Messenger \textit{et al.} \cite{PhysRevX.4.041004}  use the locations of two frequencies of the postmerger signal. They do not use the decay timescales. 
The  ratio of the two frequencies is dimensionless and depends on the total mass of the binary neutron stars. However, this ratio is not sensitive to the total mass.  In this situation, it is hard to extract the redshift from the postmerger signal alone. Fig. 4 of \cite{PhysRevX.4.041004} shows the joint posterior distributions of the redshift and total mass. The blue and red contours represent the posterior contributions from these two frequencies. Using only the contributions from these two frequencies, the redshift cannot be determined accurately. Thus, they need to combine the postmerger signal with the inspiral signal, using the inspiral signal to determine the redshifted total mass. The tidal information in the inspiral signal is ignored in \cite{PhysRevX.4.041004}.

The Fisher matrix can be used to estimate errors in the measurement of the redshift and other parameters. For any two given signals $h_1(t)$ and $h_2(t)$, the inner product is given by \cite{PhysRevD.49.2658}
\begin{equation}
(h_1|h_2) = 4 \Re \int_0^\infty \frac{\tilde{h}_1(f)\tilde{h}_2^*(f)}{S_n(f)} df
\end{equation}
where $\tilde{h}_1(f)$ and $\tilde{h}_2(f)$ are the Fourier transforms of $h_1(t)$ and $h_2(t)$; $S_n(f)$ is the  noise power spectral density (PSD).
If the waveform $h(t)$ depends on a set of parameters $\theta_i$, the Fisher matrix is given by \cite{PhysRevD.49.2658}
\begin{equation}\label{Fisher}
\Gamma_{ij} \equiv \left(\frac{\partial h}{\partial \theta_i}|\frac{\partial h}{\partial \theta_j}\right).
\end{equation}
In the large signal-to-noise ratio (SNR) regime, the expected errors $\delta \theta_i$ in the measurement of these parameters are \cite{PhysRevD.49.2658}
\begin{equation}
    \delta\theta_i=\sqrt{\Gamma^{-1}_{ii}},
\end{equation}
where $\Gamma^{-1}$ is the inverse of the Fisher matrix.
For the postmerger waveform \eqref{red postmerger}, we use the 10 parameters 
\begin{equation}\label{10para}
    \theta_i= (B_k,\phi_k, M, z_0), \quad \text{for $k$ = peak, spiral, $2\pm 0$}
\end{equation}
where the amplitude parameters $B_k\equiv \sqrt{\frac{5}{4\pi}}\frac{\bar{M}}{D_L} \mathcal{N}  A_k$. To calculate the Fisher matrix, in the postmerger waveform, the redshifted timescales $\bar{\tau}_k$ (for $k$ = peak, spiral, $2\pm 0$), the redshifted frequencies $\bar{f}_\mathrm{peak,0}$, $\bar{f}_\mathrm{spiral}$, $\bar{f}_{2\pm 0}$, and the parameters $\bar{\zeta}_\mathrm{drift}$ and $\bar{u}_*$ are expressed as functions of the source frame total mass $M$ and the redshift $z_0$. For simplicity, we set the antenna pattern functions to $F_+=1$ and $F_\times=0$. We consider measuring the redshift by two detectors,  the Neutron star Extreme Matter Observatory (NEMO) and the Einstein Telescope (ET) and set $S_n(f)$ to the noise PSDs of these two detectors \cite{etdpsd,nemopsd}.

\begin{figure}[!htbp]
\begin{center}
 \subfigure{\includegraphics[width=7cm]{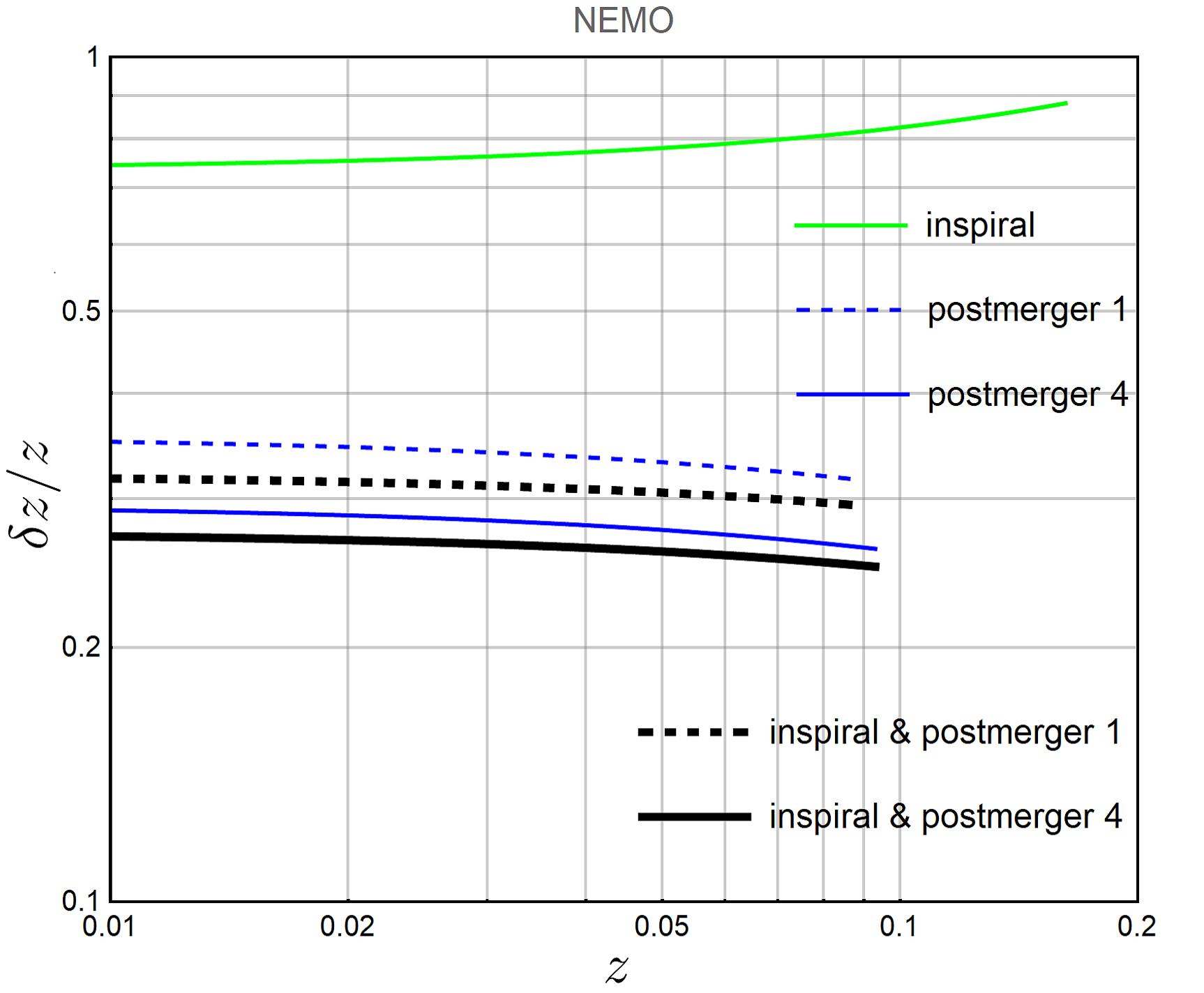}} \qquad
 \subfigure{\includegraphics[width=7cm]{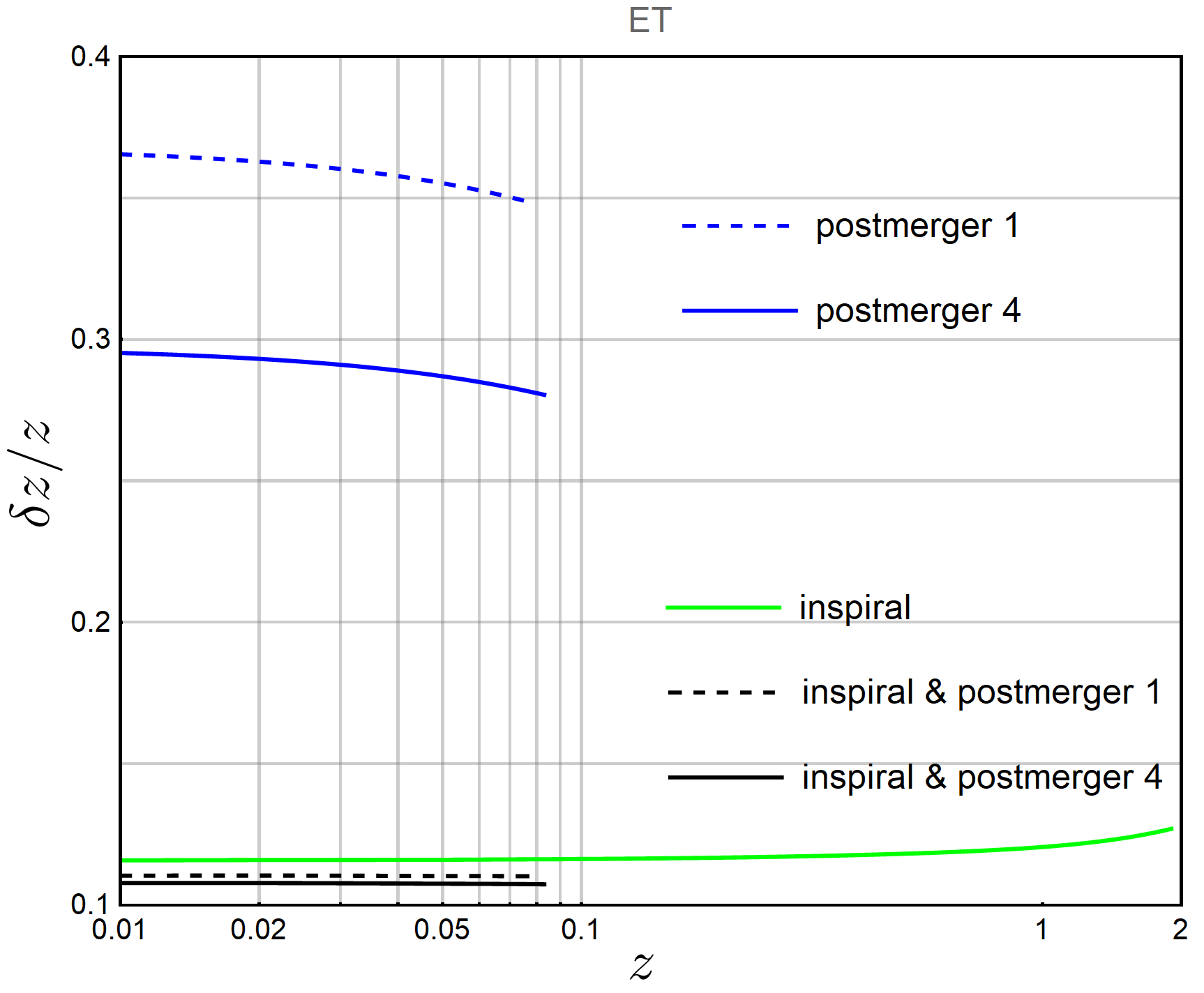}}
\caption{Fractional uncertainties in the redshift for binary neutron stars at different redshifts observed by the NEMO (left) and by the Einstein Telescope (right). We have considered extracting the redshift only from the postmerger phase (blue), only from the inspiral phase (green), and combining these two phases (black). The labels `postmerger 1' and `postmerger 4' denote the detection of one and four postmerger signal components, respectively. The equation of state MPA1 is used.
}\label{fig:red}
\end{center}
\end{figure}

The expected fractional uncertainties in the redshift $\delta z_0/z_0$, obtained by observing the postmerger signal with four components, are shown by the blue solid lines in Fig. \ref{fig:red}. For comparison, we also calculate the fractional redshift uncertainties $\delta z_c/z_c$ obtained by observing the inspiral signal following Messenger and Read's proposal \cite{PhysRevLett.108.091101}. We use the same inspiral waveform as that in \cite{PhysRevLett.108.091101} and the same parameter set $\theta_i$, while we use the equation of state MPA1 and the noise PSDs of two detectors. The result of $\delta z_c/z_c$ is shown by the green solid lines in Fig. \ref{fig:red}. The detectors can observe both the inspiral and postmerger signals, and the redshift uncertainty can be further reduced. We estimate the total redshift uncertainty as \cite{PhysRevD.101.084053}
\begin{equation}
    \left(\frac{\delta z}{z}\right)^{-2}=\left(\frac{\delta z_0}{z_0}\right)^{-2}+\left(\frac{\delta z_c}{z_c}\right)^{-2}.
\end{equation}
The total fractional uncertainties $\delta z/z$ is shown by black solid lines in Fig. \ref{fig:red}.
It is possible that only the dominant $f_\mathrm{peak}$ component of the postmerger signal is detected. In this case, to calculate the Fisher matrix, we set $k=\text{peak}$ in the parameter set \eqref{10para} and delete the three secondary components (spiral, $2\pm 0$) in the waveform \eqref{red postmerger}. The fractional redshift uncertainties, in case that only the $f_\mathrm{peak}$ component of the postmerger signal is detected, is show by dashed lines in Fig. \ref{fig:red}.
In all cases, the source frame masses of the component neutron stars are 1.4 $M_\odot$.

For a SNR threshold of 8, the detection horizon redshifts of the four-component postmerger signal, the one-component postmerger signal, and the inspiral signal  of binary neutron stars by using the NEMO or the Einstein Telescope are listed in Table \ref{horizon}.

From the left panel of Fig. \ref{fig:red}, it can be seen that, considering the NEMO  detecting the four-component postmerger signal, the redshift can be measured to better than $30\%$ accuracy for sources at $0.01<z<0.09$. If only the dominant $f_\mathrm{peak}$ component is detected, the accuracy worsens to $32\sim35\%$ for sources at $0.01<z<0.08$. For the detection of the inspiral signal, the fractional uncertainties are greater than $70\%$ for sources at $0.01<z<0.16$. Combining with the inspiral signal can reduce the redshift uncertainties obtained by observing the postmerger signal, while the reduction in the fractional uncertainties is about  $1\sim 3\%$. That is, for the NEMO, the primary contribution to the accuracy of the redshift extraction is from the detection of the $f_\mathrm{peak}$ postmerger signal, and the secondary postmerger components and the inspiral signal can moderately improve the accuracy.

Comparing the two detectors for the observation of the postmerger signal, the redshift fractional uncertainties  measured by the Einstein Telescope are larger than those measured by the NEMO, but in both the one- and four-component cases the difference is less than $4\%$. For the detection of the inspiral signal by the Einstein Telescope, the redshift fractional uncertainties are about $12\%$ for sources at $0.01<z<0.08$. Combining with the detection of the postmerger signal, the redshift fractional uncertainties can be reduced to $11\%$ for the same sources. That is, for the Einstein Telescope, the primary contribution to the accuracy of the redshift extraction is from the detection of the inspiral signal, and the postmerger signal can slightly improve the accuracy.
The  discrepancy between the two detectors' measurement capabilities stems from the fact that the noise PSD of the Einstein Telescope is comparable to NEMO's at 2 kHz, but more than an order of magnitude lower at 100 Hz.

Regarding the observation of the postmerger signal, while the redshift fractional uncertainties $\delta z/z$ decrease with increasing redshift, the redshift uncertainties $\delta z$ exhibit the opposite trend. Naively one would expect $\delta z$ to grow faster. However, there are three effects that suppress the increase. The redshifted mass $M(1+z_0)$ becomes larger, enhancing the gravitational wave amplitude; the decay timescales $\tau_i(1+z_0)$ become longer and can  help to increase the SNR; the waveforms are redshifted to lower frequency bands where the detectors are more sensitive.

\begin{table}[h]
\caption{Detection horizon redshifts of the four-component postmerger signal, the one-component postmerger signal, and the inspiral signal  of binary neutron stars by using the  NEMO and the Einstein Telescope}\label{horizon}
\begin{tabular}{lcc}
\hline\hline
  & NEMO & ET  \\ 
\hline

postmerger 4 & 0.093 & 0.083\\
postmerger 1 & 0.086 & 0.077\\
inspiral & 0.16 & 2.0  \\
\hline\hline
\end{tabular}
\end{table}

We have tried to improve the method proposed by Messenger and Read for extracting the redshift from the inspiral signal. Apart from the adiabatic tidal effect  they studied, there are dynamical tidal effects during the inspiral of the binary neutron stars. The tidal interaction can  excite the fundamental oscillation modes of neutron stars. Incorporating the dynamical tidal effect into the inspiral waveforms, the waveforms depend on the fundamental-mode frequencies $f_2^i~(i=1,2)$ \cite{PhysRevD.100.021501}. Given the equation of state of the neutron star, the dimensionless parameter $m_i\cdot f_2^i$ is a function of the neutron star mass $m_i$, which in principle can help to break the mass-redshift degeneracy. Considering the detection of the inspiral signal by the Einstein Telescope and using the equation of state MPA1, the inclusion of the dynamical tidal effect can marginally reduce the redshift fractional uncertainties, and the reduction is less than $0.2\%$ for binary neutron stars at $0.01<z<2$.

\subsection{Redshift drift effect}
We proceed to study the redshift drift effect. The procedures to obtain the waveforms incorporating the redshift drift effect, Eq. \eqref{wApsi}, apply to any isolated gravitational wave sources in the universe, such as the compact binary, the rotating neutron star, and the supernova. We focus on the inspiralling compact binary and derive the phase correction due to the redshift drift.

From Eqs. \eqref{wApsi} and \eqref{newhp}, after incorporating the redshift drift effect, the Newtonian order waveform of the inspiralling compact binary becomes
\begin{equation}
    h_+(t)=\frac{4 {\bar{M}_{c}}}{D_L} \left(\frac{5\bar{M}_{c}}{256[t_c-t+X(z_c)(t-t_c)^2]}\right)^{1/4} \frac{1+\cos^2\theta}{2}\cos[2\Phi(t-t_c-X(z_c)(t-t_c)^2, \bar{M}_{c})],
\end{equation}

The Fourier transform of the above waveform is
\begin{align}
    \tilde{h}_+(f)=&\int_{-\infty}^{+\infty}dt~e^{i 2\pi ft}~h_+(t) \nonumber \\
                  =&\frac{4 {\bar{M}_{c}}}{D_L} \frac{1+\cos^2\theta}{2} \int_{-\infty}^{+\infty}dt~e^{i 2\pi ft}\left(\frac{5\bar{M}_{c}}{256[t_c-t+X(z_c)(t-t_c)^2]}\right)^{1/4} \cos[2\Phi(t-t_c-X(z_c)(t-t_c)^2, \bar{M}_{c})] \nonumber \\
                  =&\frac{4 {\bar{M}_{c}}}{D_L} \frac{1+\cos^2\theta}{2} \int_{-\infty}^{+\infty}dt~e^{i 2\pi f[t+X(z_c)(t-t_c)^2]}\left(\frac{5\bar{M}_{c}}{256(t_c-t)}\right)^{1/4} \cos[2\Phi(t-t_c, \bar{M}_{c})] \nonumber \\
\end{align}
To obtain the last equality, we use the change of variable $t'=t-X(z_c)(t-t_c)^2$ and then discard the prime. We also use the approximation $\frac{dt'}{dt}=1$, since we focus on the phase correction to the waveform due to the redshift drift effect. The above integral can be computed via the stationary phase approximation. We rewrite it as
\begin{align}\label{spa}
    \int_{-\infty}^{+\infty}dt~e^{i 2\pi fX(z_c)(t-t_c)^2}\left(\frac{5\bar{M}_{c}}{256(t_c-t)}\right)^{1/4}\frac12 \left(e^{i[2\Phi(t-t_c, \bar{M}_{c})+2\pi f t]}+e^{i[-2\Phi(t-t_c, \bar{M}_{c})+2\pi f t]}\right)
\end{align}
For $f>0$, the first term always oscillate fast and integrates to a small value. The dominant contribution is from the second term and its stationary point $t_*$ is given by \cite{maggiore2008gravitational}
\begin{equation}
    \dot{\Phi}(t_*-t_c, \bar{M}_{c})=\pi f, \quad t_*-t_c=-\frac{5}{256}\bar{M}_c^{-5/3}(\pi f)^{-8/3}.
\end{equation}
Here, the overhead dot denotes $\frac{d}{dt}$.
The integral can be approximated by
\begin{align}
    e^{i 2\pi fX(z_c)(t_*-t_c)^2}\left(\frac{5\bar{M}_{c}}{256(t_c-t_*)}\right)^{1/4}\frac12 e^{i[-2\Phi(t_*-t_c, \bar{M}_{c})+2\pi f t_*]}\sqrt{\frac{\pi}{\ddot{\Phi}(t_*-t_c, \bar{M}_{c})}}.
\end{align}
The factor $e^{i 2\pi fX(z_c)(t_*-t_c)^2}\equiv e^{i\Psi_\mathrm{acc}(f)}$ is the phase correction due to the redshift drift (or the cosmic acceleration), where
\begin{equation}\label{psiacc}
    \Psi_\mathrm{acc}(f)=+\frac{25}{32768}X(z_c)\bar{M}_c^{-10/3}(\pi f)^{-13/3}.
\end{equation}
Previous studies yield  $-\Psi_\mathrm{acc}(f)$ for the phase correction\footnote{Eq. (2.2) in \cite{Yagi_2012} will yield $\frac{d^2t}{dt_e^2}=-g(z_c)$, which contradicts the definition of $g(z)$. Our $t_r$ here corresponds to $t_e$ in \cite{Yagi_2012}. This explains the sign error of  $\Psi_\mathrm{acc}$ in \cite{Yagi_2012}.  Thus, Eq. (2.2) in \cite{Yagi_2012} should be corrected to $\Delta t=\Delta T-X(z_c)\Delta T^2$. }. See Eq. (2) in the original paper \cite{PhysRevLett.87.221103}, Eq.(2.8) in \cite{Yagi_2012}, and Eq. (4) in a recent paper \cite{Liu_2024}. 
This sign error would lead to the interpretation of the observed cosmic acceleration as a cosmic deceleration.
We provide a different way to derive $\Psi_\mathrm{acc}$ in Appendix \ref{sec:psi_acc}.

There is an intuitive explanation of the plus sign in Eq. \eqref{psiacc}. 
When a waveform $h(t)$ is shifted to the right (delayed) by $\tau$ seconds in the time domain, it becomes $h(t-\tau)$. The Fourier transform of the shifted waveform is $\int dt~ e^{i 2\pi f t} h(t-\tau)=e^{i2\pi f\tau}\int dt ~e^{i 2\pi f t} h(t)$. This shows that a rightward time shift introduces a positive phase correction $2\pi f\tau$ in the frequency domain. From Eq. \eqref{wApsi}, the redshift drift effect can be seen as a rightward shift in the time domain, and the induced phase correction in the frequency domain should be positive.

The redshift drift effect can also be studied using the time domain waveform \eqref{wApsi}. The phase difference due to this effect is given by 
\begin{equation}\label{phase shift}
    \psi_+^{(L)}(\frac{\Delta t-X(z_0)\Delta t^2}{M(1+z_0)},\Pi)-\psi_+^{(L)}(\frac{\Delta t}{M(1+z_0)},\Pi)=-X(z_0)\Delta t^2\omega_o.
\end{equation}
Given the redshift $z_0$ of the source, a  longer observation time $\Delta t$ or a higher frequency $\omega_o$  can lead to a  larger phase difference. The ringdown signal of binary black holes is too short. The continuous gravitational waves emitted by a rotating neutron star last long enough \cite{2503.03748}. 
The SNR of a DECi-hertz Interferometer Gravitational wave Observatory (DECIGO) five-year observation of the continuous waves is of order $10^{-6}$ when the ellipticity of the neutron star is $10^{-3}$ and the source redshift is 0.1 at a frequency of 1 Hz. This means that the redshift drift effect cannot be detected by observing the rotating neutron star with DECIGO.

\section{Conclusions and Discussions}\label{condis}
We have studied the redshift effect and the redshift drift effect on gravitational waveforms due to the cosmic expansion. Using matched asymptotic expansions, we demonstrate the procedures to incorporate these two effects into the waveforms, and these procedures apply to any isolated gravitational wave sources in the universe. As an application, we show that  dimensionless parameters which depend on the mass parameter can be used to break the mass-redshift degeneracy. There are several dimensionless parameters in the postmerger waveforms of binary neutron stars that satisfy this condition, such as the quality factors of different frequency components. Observing the postmerger signal by the NEMO or the Einstein Telescope, the redshift can be determined to an accuracy $\sim 30\%$ for sources at $0.01<z\lesssim0.09$. In comparison, the quality factors of the ringdown signal of the Kerr black hole is independent of the black hole mass \cite{Maggiore2018}, and the redshift of the binary black holes cannot be extract in this way. Essentially, each mass-dependent dimensionless parameter corresponds to a length scale of the source which is not proportional to the mass. In addition, we rederive the phase correction due to the redshift drift effect to the inspiralling waveform and correct the sign error in previous studies. 
We use the toy model of scalar waves to demonstrate the method of matched asymptotic expansions. The phase evolution of scalar waves contains the redshift effect and the redshift drift effect, similar to gravitational waves. The study can be extended in several ways.

We have ignored the logarithms in the local wave zone waveforms \eqref{general_local}, which can be eliminated by the Bondi-type coordinates \cite{Blanchet2024}. To deal with the logarithms, it is necessary to use the Bondi-Sachs formalism to study the wave propagation in the local and distant wave zones. Since this formalism is only available for some special cases of the FLRW metric \cite{PhysRevD.108.064039,PhysRevD.102.104043}, further development of this formalism is desirable.

While the proof of concept analysis shows that the source redshift can be extracted by observing the postmerger signal, the fitting formulae of the parameters in the postmerger waveforms \eqref{analytic postmerger} are oversimplified. Future numerical simulations will improve the postmerger waveforms and these fitting formulae \cite{PhysRevD.104.042001,Ecker2025}, and it is important to perform the analysis based on the improved results.

There are other phenomena in binary systems related to a dimensionless parameter that depends on the mass parameter, such as the mass transfer effect \cite{PhysRevD.111.043049,Linial_2024}. The mass transfer rate $\frac{dm}{dt}$ is dimensionless and depends on the mass $m$. It is interesting to study the ability to extract the source redshift using the mass transfer effect.

The phase correction \eqref{psiacc} due to the redshift drift effect  is applicable to binary stars on a quasicircular orbit. 
Previous studies \cite{PhysRevLett.87.221103,Liu_2024} have focused on binary neutron stars. It is possible to study the redshift drift effect in other binary systems, such as binary black holes \cite{2024NatAs...8.1321S}. It would also be of interest to study binary systems on eccentric orbits.

Our analysis considers only cosmological redshift, excluding
the Doppler redshift from peculiar velocities relative to the Hubble flow \cite{PhysRevD.107.043027} and the gravitational redshift due to the matter distribution near the wave source \cite{PhysRevD.95.044029}. Future work could include these effects to generalize our results.

The method of asymptotic matching is not indispensable. The framework of Direct Integration of the Relaxed Einstein equations (DIRE) \cite{Poisson_Will_2014,PhysRevD.111.104034}, used to derive gravitational waveforms, is equivalent to the multipolar post-Minkowskian approach \cite{Blanchet2024}. While asymptotic matching constitutes a critical step in the multipolar post-Minkowskian formalism, this technique is not employed within the DIRE framework. 
As shown by Bender and Orszag, a boundary-layer problem can  be solved by either  asymptotic matching or multiple-scale analysis (see p. 559 in \cite{bender1999advanced}). This suggests that the problem addressed in this paper may  be solvable through alternative methods.

\begin{acknowledgments}
We thank Yan Wang and Tao Zhu for helpful discussions. T.L. is supported by the China Postdoctoral Science Foundation Grant No. 2024M760692. W.-F.F. is supported by the National Natural
Science Foundation of China under Grant No. 12447109.   This work is supported in part by the National Key Research and Development Program of China Grant No. 2020YFC2201501, in part by
the National Natural Science Foundation of China under Grant No. 12475067 and No. 12235019.
\end{acknowledgments}

\appendix

\section{Waves in a 1+1 dimensional universe}\label{toy}
To pedagogically illustrate the method of matched asymptotic expansions, we study  scalar waves $\varphi(t,x)$ in a $1+1$ dimensional universe, which are governed by 
\begin{equation}\label{2dwave}
\frac{1}{\sqrt{-g}}\partial_\mu(\sqrt{-g}g^{\mu\nu}\partial_\nu \varphi)=\cos(\omega t)\delta(x)\Theta(t).
\end{equation}
Here, $\delta$ denotes the Dirac delta function and $\Theta$ is the Heaviside function. $g=\det(g_{\mu\nu})$.
In this section, the spacetime metric is given by
\begin{equation}
 g_{\mu\nu}dx^\mu dx^\nu = -dt^2+a^2(t)dx^2
 \end{equation} 
where $x\in (-\infty,+\infty)$ and $t\in (-\infty,+\infty)$. This is the two dimensional  FLRW metric and $a(t)$ is the scale factor. For simplicity, there is no potential term in the scalar wave equation \eqref{2dwave}. The right hand side of Eq. \eqref{2dwave} is the source term. We assume that the behavior of the source and the scalar wave does not influence the spacetime metric. The source  starts to oscillate at $t=0$ with a constant frequency $\omega$ at the space origin $x=0$. We also assume that $\omega\gg H $. Here, $H\equiv\frac{\dot{a}}{a}$ is the Hubble parameter and  $\dot{a}\equiv \frac{d a}{dt}$. This means that the  oscillation period of the source is much smaller than the typical time scale of the universe.  The initial condition is
\begin{equation}\label{initial}
\varphi = 0 = \frac{\partial \varphi}{\partial t} \quad \text{for}  \quad  t<0.
\end{equation}
We first solve this wave equation approximately using matched asymptotic expansions, and then compare the approximate solution with the exact solution.

\subsection{Solution by matched asymptotic expansions}
The technique of matched asymptotic expansions is a  powerful and general way of finding approximate solutions. 
When there are different subregions in which the equations have rather different behaviors, the region, in which the problem is posed, is broken into a sequence of two or more overlapping subregions. Then, in each subregion an approximation to the solution of the differential equation is obtained and is valid in that subregion. Finally, the matching is done by requiring that the approximations have the same functional form on the overlap of each pair of subregions.

To solve the wave equation, we divide the space into two regions: the wave-generation zone $0\leq |x|\ll \frac{1}{\dot{a}}\big|_{t=0}$ and the wave-propagation zone $|x|>0$. 

Since the size of  the wave-generation zone is much smaller than the typical scale of the universe, the scale factor $a(t)$ can be approximated as a constant in this region $a(t)\simeq a(0) \equiv a_e$ \footnote{The subscript e stands for emission.}, the metric can be approximated by a two-dimensional Minkowski metric, and the wave equation \eqref{2dwave} becomes
\begin{equation}\label{2dflat}
(-a_e^2~\partial_t^2 +\partial_x^2)\varphi =a_e^2 ~\cos(\omega t)\delta(x)\Theta(t).
\end{equation}
The Green's function of the wave equation 
\begin{equation}
(-\frac{1}{v^2}\partial_t^2+\partial_x^2)G(x,x';t,t')=\delta(x-x')\delta(t-t')
\end{equation}
with initial conditions
\begin{equation}
G = 0 = \partial_t G \quad \text{for}  \quad  t<t'
\end{equation}
is given by (Eq. (7.3.16) in  \cite{Morse1953})
\begin{equation}\label{green}
G(x,x';t,t') = -\frac{v}{2}\Theta(t-t'-\frac{|x-x'|}{v})
\end{equation}
where $v$ is the wave velocity. Therefore, the solution to Eq. \eqref{2dflat} is 

\begin{align}
\begin{split}\label{gen-sol}
\varphi(t,x)&=\int dt'dx' ~(-\frac{1}{2a_e})\Theta(t-t'-a_e|x-x'|)~a_e^2 \cos(\omega t') \delta(x')\Theta(t')\\
&=
\begin{cases}
-\frac{a_e}{2\omega}\sin[\omega\cdot(t-a_e|x|)], \quad  &t\geq a_e|x|\\
0, &t< a_e|x|
\end{cases}
\end{split}
\end{align}

The above solution applies only in the wave-generation zone $0\leq |x|\ll \frac{1}{\dot{a}}\big|_{t=0}$; we proceed to solve Eq. \eqref{2dwave} in the wave-propagation zone $|x|>0$. Since $\omega\gg H$, the wavelength is much shorter than the typical length scale of the universe. The geometric optics approximation can be used to solve the wave propagation problem. We set the ansatz
\begin{equation}
\varphi = \Re[A e^{i\psi}]
\end{equation}
where $\Re$ denotes the real part of the argument.
The phase $\psi$ varies fast and the amplitude $A$ varies slowly
\begin{equation}
|\partial \psi|\gg |\partial A|.
\end{equation}
Substituting the ansatz into Eq. \eqref{2dwave}, in the wave-propagation zone, we have
\begin{equation}
k^\mu k_\mu A +2i k^\mu\nabla_\mu A +i\nabla^\mu k_\mu A -\nabla^\mu \nabla_\mu A=0,
\end{equation}
where 
\begin{equation}\label{kmu}
k_\mu \equiv -\partial_\mu \psi
\end{equation}
is the wave vector and $\nabla_\mu$ is the covariant derivative satisfying $\nabla_\mu g_{\alpha\beta}=0$.
Now sort this propagation equation according to the power of the wave vector.

\begin{figure}[!htbp]
\begin{center}
\includegraphics[width=8cm]{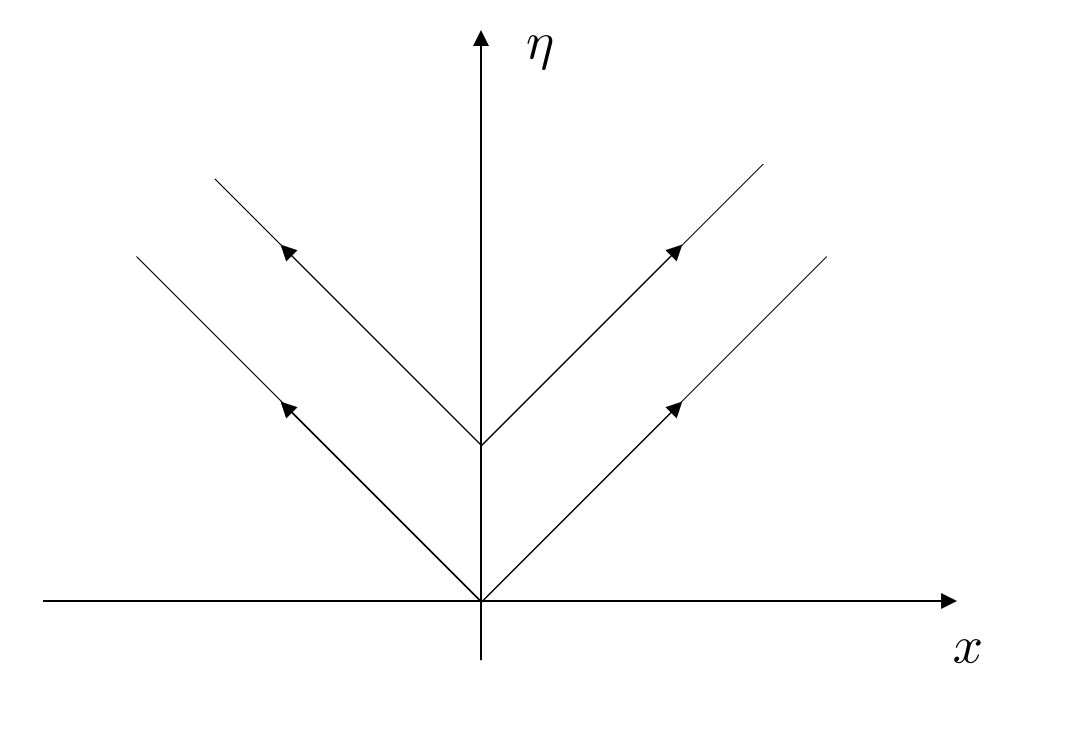}
\caption{Each null ray is specified by its conformal emission time $\eta - |x|$ and the propagation direction (left or right).
}\label{ray}
\end{center}
\end{figure}

To the leading order,
\begin{equation}\label{2deikonal}
k^\mu k_\mu =0.
\end{equation}
To the next-to-leading order,
\begin{equation}
k^\mu\nabla_\mu A +\frac12 A\nabla^\mu k_\mu =0.
\end{equation}
From Eqs. \eqref{kmu} and \eqref{2deikonal}, we have
\begin{equation}
k_\mu \nabla^\mu k^\nu =0
\end{equation}
and
\begin{equation}
k^\mu \partial_\mu \psi =0.
\end{equation}
The scalar waves propagate along null geodesics and the phase $\psi$ is constant along each null geodesic. The expression $\nabla^\mu k_\mu$ denotes the change of the cross-sectional area of the congruence of the null geodesics \cite{Poisson2004}, and the cross-sectional area is always zero since there is only one spatial dimension. Therefore, $\nabla^\mu k_\mu=0$ and 
\begin{equation}
k^\mu \partial_\mu A =0.
\end{equation}
The amplitude $A$ is also constant along each null geodesic.
It can be seen from Fig. \ref{ray} that each null geodesic is specified by its emission time (i.e. the retarded time $t_r$) and its propagation direction. In addition, the wave equation \eqref{2dwave} and its initial conditions \eqref{initial} are invariant under the transformation $x\to -x$. Therefore, the phase $\psi$ and the amplitude $A$ as functions of two variables $(t,x)$ depends only on their combination $t_r$, 
\begin{equation}\label{pro-sol}
\psi(t,x)=\psi(t_r),\quad A(t,x)=A(t_r).
\end{equation}
Now, the wave-propagation problem has been solved. The explicit expressions of $\psi$ and $A$ should be determined by the solution to the wave-generation problem.

To express the retarded time $t_r$ explicitly, we introduce the conformal time $\eta$. Its relation with the physical time $t$ is given by
\begin{equation}
t=\int_0^\eta a(\eta') d\eta'.
\end{equation}
and the metric becomes
\begin{equation}
ds^2 = a^2(\eta)(-d\eta^2+dx^2).
\end{equation}

It is obvious that  $\eta - |x|$, the conformal emission time, is constant along each null geodesic. The emission time, i.e. the retarded time, takes the form
\begin{equation}
t_r = \int_0^{\eta - |x|} a(\eta') d\eta'.
\end{equation}

We will now move on to the matching procedure in the overlapping region
$0 < |x|\ll \frac{1}{\dot{a}}\big|_{t=0}$. 
In this region, both the the wave-generation solution, Eq. \eqref{gen-sol}, and the wave-propagation solution, Eq. \eqref{pro-sol}, are valid and should take the same form.
We rewrite Eq. \eqref{gen-sol} in the following form
\begin{equation}\label{appro-gen}
\varphi= \Re[-\frac{a_e}{2\omega}e^{i[\omega\cdot(t-a_e|x|)-\frac{\pi}{2}]}]\Theta(t-a_e|x|).
\end{equation}
In the overlapping region, the retarded time can be approximated as
\begin{equation}
t_r = \int_0^\eta a(\eta') d\eta' -\int_{\eta-|x|}^\eta a(\eta')d\eta'\simeq t-a_e|x|.
\end{equation}
As a consequence, Eq. \eqref{pro-sol} can be approximate as
\begin{equation}
\psi(t,x) \simeq \psi(t-a_e|x|),\quad A(t,x) \simeq A(t-a_e|x|).
\end{equation}
Matching the above with Eq. \eqref{appro-gen}, we have
\begin{equation}
\psi(t-a_e|x|)=\omega\cdot(t-a_e|x|)-\frac{\pi}{2},\quad A(t-a_e|x|)=-\frac{a_e}{2\omega}\Theta(t-a_e|x|)
\end{equation}
Therefore, in the wave-propagation zone $|x|>0$, we have
\begin{equation}\label{maephi}
\varphi =-\frac{a_e}{2\omega} \sin(\omega t_r) \Theta(t_r).
\end{equation}
Compared with the solution \eqref{gen-sol} on the flat background, it can be seen that  replacing $t-a_e |x|$ in Eq. \eqref{gen-sol} with $t_r$ can yield the above result, which reflects the influence of cosmic evolution on the waves.
The observed frequency is
\begin{equation}\label{omegao}
\omega_o = \frac{d(\omega t_r)}{dt}= \omega\frac{a(t_r)}{a(t)}.
\end{equation}
The time evolution of the observed frequency has some interesting properties that will be useful in the following sections.

\subsection{Comparison with the exact solution}
\begin{figure}[!htbp]
\begin{center}
\includegraphics[width=8cm]{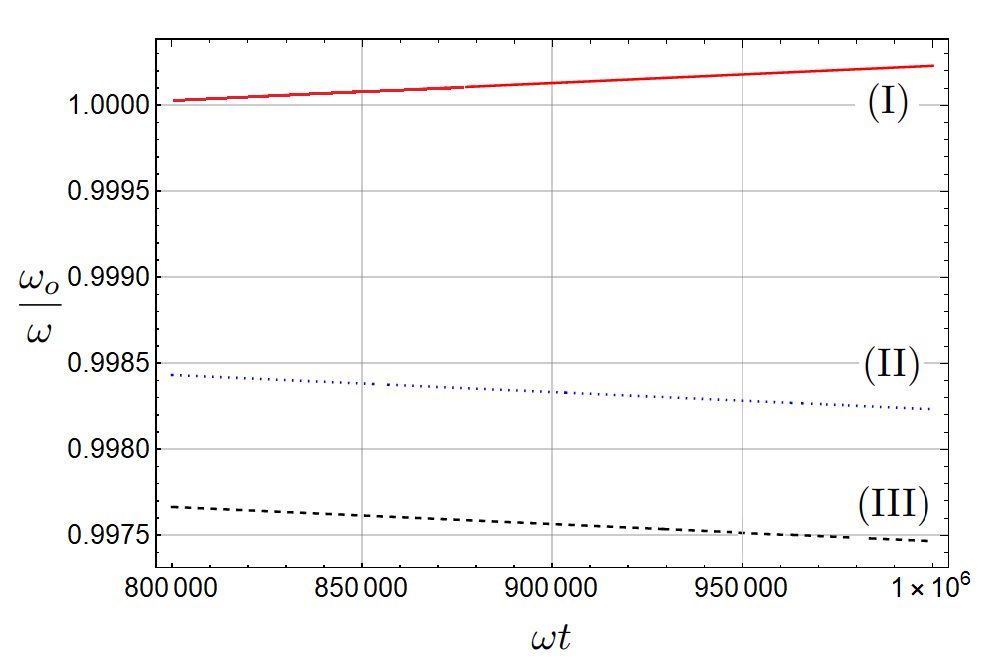}
\caption{Influence of cosmic evolution on the time evolution of the observed frequency $\omega_o$. The scale factor is parametrized by $a_1$ and $a_2$, $a(t)=1+a_1 t+\frac12 a_2 t^2$.  Different evolution behaviors of $\omega_o$ are due to different choices of parameters $(a_1,a_2)$. Parameter values are for illustrative purposes only. For the red solid line  (\uppercase\expandafter{\romannumeral1}), $(a_1,a_2)=(\frac{\omega}{1.3\times 10^{6}},-\frac{\omega^2}{10^{12}})$. For the blue dotted line  (\uppercase\expandafter{\romannumeral2}), $(a_1,a_2)=(\frac{\omega}{1.3\times 10^{6}}, \frac{\omega^2}{10^{12}})$. For the black dashed line  (\uppercase\expandafter{\romannumeral3}), $(a_1,a_2)=(\frac{2\omega}{1.3\times 10^{6}},\frac{\omega^2}{10^{12}})$.  For all the three cases, the observer is located at $x=\frac{1000}{\omega}$.
}\label{omega}
\end{center}
\end{figure}

We proceed to compare the above solution obtained by matched asymptotic expansions with the exact solution. Using the coordinates $(\eta,x)$,
the wave equation \eqref{2dwave} becomes
\begin{equation}
(-\partial_\eta^2 +\partial_x^2)\varphi = a^2\cos(\omega t) \delta (x) \Theta (t).
\end{equation}
Using the Green's function \eqref{green}, we have 
\begin{equation}
\varphi =\left [ -\frac{a(t_r)}{2\omega} \sin(\omega t_r) +\frac12 \int_0^{t_r} \frac{\dot{a}}{\omega}\sin(\omega t) dt \right ] \Theta(t_r).
\end{equation}
It can be seen that the second term is of order $\mathcal{O}(\frac{H}{\omega})$
relative to the first term. Comparing the approximate solution \eqref{maephi} with the first term, the approximate solution has the exact phase, but ignores the influence of the cosmic expansion on the amplitude.
Since the emission frequency $\omega$ is constant, the time evolution of the observed frequency $\omega_o$ is determined by the time evolution of the scale factor $a(t)$. As illustrative examples, Fig. \ref{omega} shows the time evolution of the observed frequency $\omega_o$ for different evolutionary behaviors of the scale factor $a(t)$. It can be seen that $\omega_o$ is redshifted and the redshift evolves with time, which is the redshift drift effect. The evolution of $\omega_o$ depends on the Hubble parameter and the deceleration parameter which correspond to the first and second time derivatives of the scale factor, respectively. If the scale factor is a linear function of time, there is no redshift drift. In the four-dimensional FLRW universe, the frequency of the gravitational waves has the similar time dependence (cf. Eq \eqref{omegao_x}). It is proposed that the  DECIGO can directly measure the redshift drift effect due to the acceleration of the universe \cite{PhysRevLett.87.221103,Yagi_2012}.

\section{Fitting results of the parameters in the secondary components of the postmerger waveform}\label{sec:fitting_secondary}
Eqs. \eqref{equations-fpeak(t)-fits_1}-\eqref{phipeak-Mtot} are the fitting formulae of the parameters of the $f_\mathrm{peak}$ component. Here are the fitting formulae of  the frequencies $f_k$, the decay timescales $\tau_k$, the dimensionless amplitudes $A_k$,  the initial phases $\phi_k$  (for $k$ = spiral, $2\pm 0$), and the normalization factor $\mathcal{N}$ \cite{Soultanis2022}.

\begin{eqnarray}
\label{fspiral of mtot}f_\mathrm{spiral} &=& +0.319~M^2-0.758~M+1.914,\\
\label{f2-0 of mtot}f_{2-0} &=& +0.236~M^2+0.167~M-0.433,\\
\label{f2+0 of mtot}f_{2+0} &=& +0.371~M+2.929,
\end{eqnarray}

\begin{eqnarray} 
\label{Tspiral of mtot}\tau_\mathrm{spiral} &=& -0.874~M^2+3.521~M-2.005, \\
\label{T2-0 of mtot}\tau_{2-0} &=& +2.057~M^2-10.804~M+14.606,\\
\label{T2+0 of mtot}\tau_{2+0} &=& +8.469~M^2-48.785~M+71.671,
\end{eqnarray}

\begin{eqnarray}
\label{Aspiral of mtot}A_\mathrm{spiral} &=& +2.649~M^2-13.580~M+17.752,\\
\label{A2-0 of mtot}A_{2-0} &=& -1.704~M^2+10.004~M-13.909,\\
\label{A2+0 of mtot}A_{2+0} &=& +0.816~M^2-3.920~M+4.734,
\end{eqnarray}

\begin{eqnarray}
\label{phispiral-Mtot}\phi_{\mathrm{spiral}} =    \left\{
\begin{array}{ll}
+17.580~M-42.199, & \mbox{for } M\leq 2.7~M_\odot \nonumber\\
+40.448~M-104.258, &\mbox{for }  M >  2.7~M_\odot  \\
\end{array} 
\right. 
\\
\\
\label{phi2-0-Mtot}\phi_{\mathrm{2-0}} =    \left\{
\begin{array}{ll}
+18.541~M-43.911, & \mbox{for } M\leq 2.7~M_\odot \nonumber\\
+43.613~M-112.705, & \mbox{for } M >  2.7~M_\odot  \\
\end{array} 
\right. 
\\
\\
\label{phi2+0-Mtot}\phi_{\mathrm{2+0}} =    \left\{
\begin{array}{ll}
+16.064~M-41.163, & \mbox{for } M\leq 2.7~M_\odot \nonumber\\
+43.309~M-115.341, & \mbox{for } M >  2.7~M_\odot  \\
\end{array} 
\right. 
\\
\end{eqnarray}

\begin{eqnarray}
\label{N of Mtot}\mathcal{N} &=& -0.485~M+2.025.
\end{eqnarray}

\begin{table}[ht]
        \begin{tabular}{l|c}
        \hline\hline
            \textrm{Symbol}&
            \textrm{Unit}
            \\
            \hline
        $M$& $M_\odot$\\
        $u_*$& $\mathrm{ms}$\\
        $\zeta_\mathrm{drift}$& $\mathrm{kHz^2}$\\
        $f_\mathrm{peak,0}$& $\mathrm{kHz}$\\
        $f_\mathrm{spiral}$& kHz\\
        $f_\mathrm{2\pm0}$& kHz\\
        $\tau_\mathrm{peak}$& ms\\
        $\tau_\mathrm{spiral}$& ms\\
        $\tau_\mathrm{2\pm0}$& ms\\
        $A_\mathrm{peak}$& dimensionless\\
        $A_\mathrm{spiral}$& dimensionless\\
        $A_\mathrm{2\pm0}$& dimensionless\\
        $\mathcal{N}$& dimensionless\\
        $\phi_\mathrm{peak}$& rad\\
        $\phi_\mathrm{spiral}$& rad\\
        $\phi_\mathrm{2\pm0}$& rad\\
        \hline\hline
        \end{tabular}
    \caption{The units of the parameters in the fitting equations \eqref{equations-fpeak(t)-fits_1}-\eqref{phipeak-Mtot}} and \eqref{fspiral of mtot}-\eqref{N of Mtot} \cite{Soultanis2022}.
    \label{Units}
\end{table}

\section{Different way to derive $\Psi_\mathrm{acc}$}\label{sec:psi_acc}
To derive the phase correction due to the redshift drift, we rewrite the integral \eqref{spa} as
\begin{align}
    \int_{-\infty}^{+\infty}dt~\left(\frac{5\bar{M}_{c}}{256(t_c-t)}\right)^{1/4}\frac12 \left(e^{i[2\Phi(t-t_c, \bar{M}_{c})+2\pi f (t+X(z_c)(t-t_c)^2)]}+e^{i[-2\Phi(t-t_c, \bar{M}_{c})+2\pi f (t+X(z_c)(t-t_c)^2)]}\right).
\end{align}
The stationary point $T_*$ of the second term is determined by
\begin{equation}
    \dot{\Phi}(T_*-t_c, \bar{M}_{c})=\pi f[1+2X(z_c)(T_*-t_c)].
\end{equation}
To solve this equation, we expand $T_*$ as
\begin{equation}
    T_*-t_c=(T_0-t_c)(1+X(z_c)T_1),
\end{equation}
where $T_0$ is the stationary point without the redshift drift effect, satisfying
\begin{equation}\label{T0}
    \dot{\Phi}(T_0-t_c, \bar{M}_{c})=\pi f,
\end{equation}
and $T_1$ is the correction of the stationary point due to the redshift drift.
Therefore, the above integral can be approximated by
\begin{align}
    \frac12 e^{i[-2\Phi(T_*-t_c, \bar{M}_{c})+2\pi f (T_*+X(z_c)(T_*-t_c)^2)]}\left(\frac{5\bar{M}_{c}}{256(t_c-T_0)}\right)^{1/4} \sqrt{\frac{\pi}{\ddot{\Phi}(T_0-t_c, \bar{M}_{c})}}, 
\end{align}
where we have discarded the amplitude correction due to the redshift drift. The phase can be expanded as
\begin{align}
    &-2\Phi(T_*-t_c, \bar{M}_{c})+2\pi f [T_*+X(z_c)(T_*-t_c)^2] \nonumber \\
   =&-2\Phi(T_0-t_c, \bar{M}_{c})-2\dot{\Phi}(T_0-t_c, \bar{M}_{c})X(z_c)(T_0-t_c)T_1+2\pi f [T_0+X(z_c)(T_0-t_c)T_1+X(z_c)(T_0-t_c)^2] \nonumber \\
   =&-2\Phi(T_0-t_c, \bar{M}_{c})+2\pi f T_0+2\pi f X(z_c)(T_0-t_c)^2. 
\end{align}
The last term in the last row is the phase correction due to the redshift drift, which is consistent with Eq. \eqref{psiacc}.
We have used Eq. \eqref{T0}  to eliminate  terms proportional to $T_1$ and discarded terms of order $\mathcal{O}(X(z_c)^2)$.


%

\end{document}